\documentclass{elsart}
\usepackage{natbib}
\usepackage{graphicx}
\begin{document}
\begin{frontmatter}
\title{Photometrical Review of Open Cluster M25}

\author[NRIAG]{A.L. Tadross\thanksref{1}},
\author[NRIAG]{A.I. Osman\thanksref{2}},
\author[Cairo]{M.A. Marie\thanksref{3}},
\author[NRIAG]{S.M. Hassan\thanksref{4}},

\thanks[1]{E-mail: altadross@mailer.scu.eun.eg; altadross@yahoo.com (A.L. Tadross), \\
Phone: +202 5560645; Fax: +202 5548020}
\thanks[2]{E-mail: amiosman@mailer.scu.eun.eg (A.I. Osman)}
\thanks[3]{E-mail: mohmarie@mailer.scu.eun.eg (M.A. Marie)}
\thanks[4]{E-mail: astro1@frcu.eun.eg (S.M. Hassan)}

\address[NRIAG]{National Research Institute of Astronomy and Geophysics,
Cairo, Egypt}
\address[Cairo]{Astronomy Department, Faculty of Science, Cairo University, Egypt}

\begin{abstract}
The young open star cluster M25 (IC 4725) is located in the
direction of the galactic center in a crowded region, near much
irregular absorption features on Sagittarius arm. This cluster has
some difficult observing problems due to its southern location.
The mass data available in the literature have been gathered to
reinvestigate this cluster using most photometric tools to
determine its main photometric parameters. This system is found to
be at a distance of 600 pc from the sun; distance of -52.82 pc
from the galactic plane; and the median age is 9.45 x 10$^{7}$ yr.
More than 220 stars with mean reddening of 0.50 mag and absorption
of 1.62 mag are found within the cluster.
\end{abstract}
\end{frontmatter}

\section*{1. Introduction}
The galactic cluster M25 is a  moderately  young  open  star
cluster (Sandage 1963 \cite{San63}). Its 2000.00 position on the
celestial sphere is $\alpha=18^{h} 31.6^{m} \ \& \
\delta=-19^{\circ} 15^{'}$ and $\ell=13.92^{\circ}  \ \&  \
b=-5.05^{\circ}$. The Palomar Observatory Sky Survey (POSS-print)
shows that the cluster M25 has several bright stars and many of
intermediate brightness superposed upon numerous faint background
stars. This system has a lot of irregular interstellar matter due
to its location near the galactic plane (Landolt 1964
\cite{Lan64}; Schmidt 1977 \cite{Sch77}). Also, there is a dark
lane of absorbing matter passing near the center of the cluster
(Serkowski 1965 \cite{Ser65}; Schmidt 1982 \cite{Sch82}). A
Cepheid Variable (U-Sgr) with a period of 6.745241$^{d}$ lies near
the center of the cluster (Wallerstein 1957 \cite{Wal57};
Shobbrook 1992 \cite{Sho92}).

Many previous studies have been performed on this cluster since
1957, e.g. Feast 1957 \cite{Fea57}; Wallerstein 1957 \cite{Wal57};
Irwin 1958 \cite{Irw58}; Sandage 1960 \cite{San60}; Johnson 1960a
\cite{Joh60a}; Wampler et al. 1961 \cite{Wam61}; Landolt 1964
\cite{Lan64}; Serkowski 1965 \cite{Ser65}; Graham 1967
\cite{Gra67}; Schmidt 1977 \cite{Sch77}; van den Bergh 1978
\cite{Van78}; Schmidt 1982 \cite{Sch82}; Stephen 1989
\cite{Ste89}; Luck et al. 2000 \cite{Luc00}; and Tadross et al.
2002 \cite{Tad02}.

It is noted that there is a deficiency of observations in U-band,
and only observations down to 15.5$^{th}$ visual magnitude are
available. This certainly will make it difficult to define the
main sequence of the cluster and consequently affect the
estimation of the main photometric parameters. So, the main data
of van den Bergh (1978)\cite{Van78}, Landolt (1964)\cite{Lan64}
and Stephen (1989)\cite{Ste89} have been used after being reduced
to van den Bergh's system in the present study.

\section*{2. Photometric analysis of M25}
\subsection*{2.1. Reddening}
The interstellar reddening across  the  cluster  has  been
determined using its color-color diagram {\it (CCD)}, as shown in
Fig 1. The curve represents the standard zero age  main sequence
{\it (ZAMS)} as taken from Schmidt-Kaler (1982)\cite{SchK82}. The
slope of the reddening line has been considered to be  0.72, as
given by Johnson \& Morgan  (1953)\cite{Joh53}. A line parallel to
the reddening line has been drawn for every  point on the {\it
CCD}. The intersection of this line with the {\it ZAMS}-curve
gives the intrinsic values of the color indices assuming that the
star lies on the main sequence. The reddening of any  star lying
below the kink of the {\it ZAMS}-curve is ambiguous and may have
two or three possible values. For this reason, the best reading
that is consistent with the cluster distance has been accepted.

{\bf \S}\ A histogram of E(B-V) has been constructed for all
members of the cluster. A range of the reddening from 0.30 mag to
0.70 mag has been obtained as shown in Fig 2. The model value for
all the cluster stars is 0.48 mag (solid line), and 0.50 mag
(dashed line) for the members with unambiguous reddening. The
maximum and minimum values of the reddening estimated by Burki's
method (1975)\cite{Bur75}, Fig 3, come out to be 0.63 mag \& 0.37
mag respectively. The difference between the maximum and minimum
reddening has found to be $\Delta$E(B-V)= 0.26 mag, which is
larger than the natural dispersion of the cluster (0.11 mag)
according to Burki's criterion. This indicates that the reddening
across the cluster is non-uniform.

{\bf \S}\ The Q-method of Johnson \& Morgan (1953)\cite{Joh53} has
been applied to estimate the maximum and minimum reddening of the
cluster. It's found E(B-V)$_{max}$= 1.20 mag \& E(B-V)$_{min}$=
0.35 mag. Also the difference between the maximum and minimum
excesses, $\Delta$E(B-V), is found to be greater than Burki's
(1975)\cite{Bur75} limit of the natural dispersion.

{\bf \S}\ From the relation between the color  excesses  and  the
distance modulii of the cluster members, Fig 4; the reddening is
found to range from 0.15 to 1.00 mag, and the mean excess changes
with the mean distance from 0.47 to 0.52 mag, Fig 5.

{\bf \S}\ The cluster reddening has been also estimated from the
spectral classifications of Wampler et al (1961)\cite{Wam61},
Feast (1957)\cite{Fea57}, Wallerstien (1957)\cite{Wal57}, and
Wallerstien (1960)\cite{Wal60} data. The mean value of the
reddening has been calculated to be 0.47 mag (18 stars), 0.45 mag
(29 stars) and 0.46 mag (8 stars) respectively. Intercomparisons
of E(B-V) values for the three investigators have been established
in the  sense of 3-color photometric data minus the spectroscopic
ones as shown in Fig 6. In this respect, we found that Wampler's
spectroscopic data is in a good agreement with the photometric
data than those of Feast and Wallerstien.

{\bf The mean reddening value of the cluster has been taken to be
0.50 mag, which is in a good agreement with those obtained above.
It coincides with Johnson \& Beckers' methods (see section 3).}

\subsection*{2.2. Interstellar absorption}
It is assumed that there is a constant ratio between the total
visual absorption A$_{v}$ and the color excess E(B-V), which
depends on the nature of  the dust grains. Blanco
(1955)\cite{Bla55} has shown that this ratio is not always
constant and it seems to be uniform except for a  very  few
anomalous directions  in  the sky; e.g., Orion nebula. Moffat \&
Schmidt-Kaler (1976)\cite{Mof76} have revised the ratio of total
visual absorption to selective color excess in  the long
wavelength band from the relation [A$_{v}$/E(B-V)={\it R} ]
assuming the constant {\it R} is 3.25 for most directions in the
sky. Applying this relation in the direction to the cluster under
investigation, the total visual absorption A$_{v}$ is found equal
1.62 mag.

\section*{3. Distance}
According to Johnson's (1960b)\cite{Joh60b} method, the reddening
can be obtained by shifting the standard {\it CCD}-curve of
Schmidt-Kaler (1982)\cite{SchK82} on to the apparent {\it
CCD}-curve of the cluster in the direction of the reddening line;
the slope of this line has been considered to be 0.72 as explained
by Johnson \& Morgan (1953)\cite{Joh53}; this yields an excess of
0.50 mag. From the coincidence of the corrected sequence of the
cluster's color-magnitude diagram, {\it CMD}-curve, with the
standard one of Schmidt-Kaler (1982)\cite{SchK82}, the apparent
distance modulus is found to be 10.55 mag. Subtracting the total
visual absorption, the true distance modulus is found to be 8.93
mag.

{\bf \S}\ Becker's (1972)\cite{Bec72} method has been applied,
where the intrinsic standard {\it CMDs} of Schmidt-Kaler
(1982)\cite{SchK82} have been superposed on the apparent ones of
the cluster [V,(B-V) \& V,(U-B)] under some conditions assumed by
Becker \& Stock (1954)\cite{Bec54}. The well known Becker's table
has been constructed, Table 1, where the apparent distance modulii
{\it V-M$_{v}$} in the first column are plotted versus {\it
$\triangle$E(U-B)$_{o-c}$} values in the last column, Fig 7. The
true apparent distance at which $\triangle$E = 0.0 is found to be
10.58 mag; it's conforming with a color excess of about 0.50 mag.
After subtracting the total visual absorption, the true distance
modulus is found to be 8.96 mag.

{\bf \S}\ A histogram of {\it V$_{o}$-M$_{v}$} has been
constructed for all the cluster members as shown in Fig 8. The
model value for the cluster is found to be 8.65 mag (solid line),
while for unambiguous members (stars which are above the kink of
the {\it CCD}-curve as shown in Fig 1), it is found to be 8.75 mag
(dashed line). Then the mean distance modulus is equal to 8.70
mag.

{\bf \S}\ From the relation between the distance modulii and
color excesses, Fig 5, we could infer that the cluster modulus has
a distance range from 7.75 mag to 10.20 mag and, as a consequence,
an average value of 8.98 mag.

{\bf \S}\ From the empirical mean evolutionary  deviation curve of
Johnson (1960b)\cite{Joh60b}, Fig 9; an estimation of the distance
modulus of 9.10 mag to the cluster could be achieved.

{\bf \S}\ Using the spectroscopic data of Wampler et al
(1961)\cite{Wam61}, Feast (1957)\cite{Fea57}, Wallerstein
(1957)\cite{Wal57}, and Wallerstein (1960)\cite{Wal60}. The mean
modulus of the cluster is found to be 8.25 mag (18 stars), 8.18
mag (29 stars), and 8.24 mag (8 stars) respectively. It is noticed
that Wampler's spectroscopic results are in a good agreement with
the 3-color photometric data than those obtained by Feast and
Wallerstein, Fig 10.

{\bf \S}\ The distance modulus of the cluster has been obtained by
fitting the {\it ZAMS}-curves of Schmidt-Kaler (1982)\cite{SchK82}
to the lower portions of the [V$_{o}$ , (B-V)$_{o}$] and [V$_{o}$
, (U-B)$_{o}$] diagrams of the cluster as shown  in Figs 11 \& 12.
The mean value of the distance modulus comes out to be 8.93 mag.

{\bf \S}\ The mean distance modulus of the cluster has been
calculated from the whole individual modulii of the members
\begin{center}
(V-Mv)$_{o}$ = 8.82 mag $\pm$ 0.006 (standard error of the mean)
\end{center}

{\bf \S}\ The classical Cepheid U-Sgr (star no 1457 in the present
work) is found to have a distance modulus ranging from 8.95 mag to
8.99 mag according to Wolfgang (1988)\cite{Wol88} and Visvanatha
(1989)\cite{Vis89} respectively. This value is in a good agreement
with the present distance of the cluster.

{\bf The assigned mean distance modulus of the cluster is then
8.89 mag. This corresponds to a distance of 600 pc to it.}

The cluster's distances from the galactic plane {\it Z} and from
the galactic center {\it R$_{gc}$} have been calculated to be
-52.82 pc and 7.92 kpc respectively.

\section*{4. Membership analysis}
The assignment of membership or non-membership based on the
photometric analysis is very difficult in this cluster because the
presence of differential reddening all over the cluster. The
uncertainty about membership increases for faint stars those are
merging with crowded galactic field. So that the discrimination of
the physical members of the cluster from the field stars could be
better achieved by studying the radial velocities and relative
proper motions.

In the present work, our assignment of membership probabilities
has been based on the position of each star in the long
wave-length and short wave-length CMDs of the cluster with the
proper motion studies of Stephen (1989)\cite{Ste89}. Also, the
distance, the color excess and the spectral type have been taken
into account in our decisions about some stars. Hence, 115 stars
are regarded as members based on their good positions in the two
CMs, distance, proper motion membership probabilities ($>$ 50\%),
color excess, and spectral type ranges.

From the studies, which have only {\it BV} observations, almost
120 stars have been added to the cluster as probable members based
on their location near the {\it ZAMS}-curve of the long
wave-length CMD, as shown in Fig 13. Also, the proper motion
membership probabilities ($>$ 50\%) have been taken into account.
Therefore, the mean distance and reddening of these stars are
found to be in a good agreement with the distance modulus and
reddening ranges of the cluster. So that, the cluster M25 may be
contain at least 230 stars as members and probable members.

{\it Moreover, the free reddened CMDs of the cluster have been
used to separate the doubt members from the normal ones as
follows:}

- According the previous studies, stars no 60, 118, \& 127 are
classified as foreground stars and stars no 11, 20, 51, 56, 73, \&
534 as background ones. In the present work, based on our
considerations, the stars no 11 \& 20 have accounted as probable
members.

- Stars no J1 \& J106 {taken from Johnson's (1960a)\cite{Joh60a}
sequence} are regarded as members because they are consistent with
our conditions for the cluster members, then they have been added
to the present sequence.

- Stephen (1989)\cite{Ste89} suggested that the star no 73 is a
cluster member with a proper motion membership probability of
94\%. While, Feast's (1957)\cite{Fea57} and Schmidt's
(1977)\cite{Sch77} studies support our suggestion to regard this
star as a non-member one.

- Stars no 19 \& 45 have the brightest magnitudes in the CMDs and
almost have the same radial velocity of the cluster (Feast
1957\cite{Fea57}, Wallerstien 1957\cite{Wal57} \& Wallerstien
1960\cite{Wal60}). So, Feast \& Evan (1967)\cite{Fea67} classified
them as variable members supported by Wampler et al
(1961)\cite{Wam61}, Schmidt (1977)\cite{Sch77} \& Schmidt
(1982)\cite{Sch82}.

- The two stars no 82 \& 885 are classified as evolved stars and
found by Schmidt (1984)\cite{Sch84} to be yellow giant cluster
members. Accordingly, it is suggested  that the  cluster M25
contains three evolved stars: the Cepheid U-Sgr; and the two
giants no 82 \& no 885.

\section*{5. The field stars}
Most stars, which haven't enough evidences supporting their
membership to the cluster have been regarded as background field
stars. This field is found to contain at least 1000 faint stars,
as shown in Fig 14. Their distribution in the long wave-length CMD
has a lower main sequence than that of the cluster members, Fig
15. The fit of the standard long wave-length CMD of Schmidt-Kaler
(1982)\cite{SchK82} with the apparent one of the cluster's field
has shown that these stars have a distance modulus of 12.30 mag
and color excess of 0.67 mag. This result is coinciding with the
fact that the color excess and absorption increase with distance.

\section*{6. Age of the cluster}
In the previous studies of M25, Wampler's (1961)\cite{Wam61}
UBV-photometry yields an age of the cluster lies between
Mermilliod's (1981)\cite{Mer81} age groups $\alpha$-persei and IC
4665, i.e. between 3.6 and 5.1 x 10$^{7}$ yr. On the other hand,
the age of the cluster has been estimated by Tosi
(1979)\cite{Tos79} as 8.9 x 10$^{7}$ yr, and by Strobel
(1989)\cite{Str89} as 10$^{8}$ yr. In the present work, the age
has been estimated using Sandage's (1957)\cite{San57} relation
({\it M$_{v}$-log t}) where the visual absolute magnitude M$_{v}$
of the turn off point of the cluster main sequence is found to lie
at (B-V)$_{t}$ = -1.2 mag; this value related to {\it log t} of
7.95, i.e. an age of  9.1 x 10$^{7}$ yr.

As an additional check, the age of the cluster has been estimated
using the isochrones curves of Barbaro et al (1969)\cite{Bar69},
which yields an age of  9.8 x 10$^{7}$ yr. Our estimation of the
age shows a range from 9.1 x 10$^{7}$ to 9.8 x 10$^{7}$ yr. This
estimation is greater than that  obtained by  Mermilliod's
(1981)\cite{Mer81}, smaller than what given by Strobel
(1989)\cite{Str89}, and almost in a good agreement with that
obtained by Tosi (1979)\cite{Tos79}.

\section*{7. Conclusions}
The mean values of reddening and distance are found to be 0.50 mag
and 600 pc respectively. The Cepheid variable U-Sgr has been used
to support our distance estimation to the cluster. The cluster's
distances from the galactic plane {\it Z} and from the galactic
center {\it R$_{gc}$} have been calculated to be -52.82 pc and
7.92 kpc respectively. The median age of the cluster has been
achieved to get an age range from 9.1 x 10$^{7}$ to 9.8 x 10$^{7}$
yr. The membership analysis has been analyzed and about 120 stars
have been added to the cluster as probable members based on their
good location in CMDs, distance modulus, color excess and the
proper motion membership probabilities ($>$ 50\%).

\section*{Acknowledgment}
We want to thank Prof. Dr. Roland Buser  (Basel Univ.,
Switzerland) for his critically reading the original task of this
work, and for his valuable discussions during his visitation to
Egypt.

\begin{table}
\caption{Becker's method for estimating the distance to M25}
\begin{tabular}{lllll}
\hline\noalign{\smallskip}V-M$_{v}$&E(B-V)$_{obs.}$&E(U-B)$_{obs.}$&E(U-B)$_{cal.}$&$\delta$
E(U-B)$_{o-c}$
\\\hline\noalign{\smallskip}
9.40&0.28&0.13&0.20&- 0.07\\
9.60&0.32&0.17&0.23&- 0.06\\
9.80&0.37&0.22&0.27&- 0.05\\
10.00&0.40&0.25&0.29&- 0.04\\
10.20&0.44&0.29&0.32&- 0.03\\
10.40&0.47&0.32&0.34&- 0.02\\
10.60&0.50&0.37&0.36&+0.01\\
10.80&0.53&0.40&0.38&+0.02\\
11.00&0.55&0.43&0.40&+0.03\\
11.20&0.58&0.45&0.42&+0.03\\
11.40&0.60&0.48&0.43&+0.05\\
\hline{\smallskip}
\end{tabular}
\end{table}

\newpage
\begin{figure}
\centerline{\includegraphics[height=22cm,width=20cm]{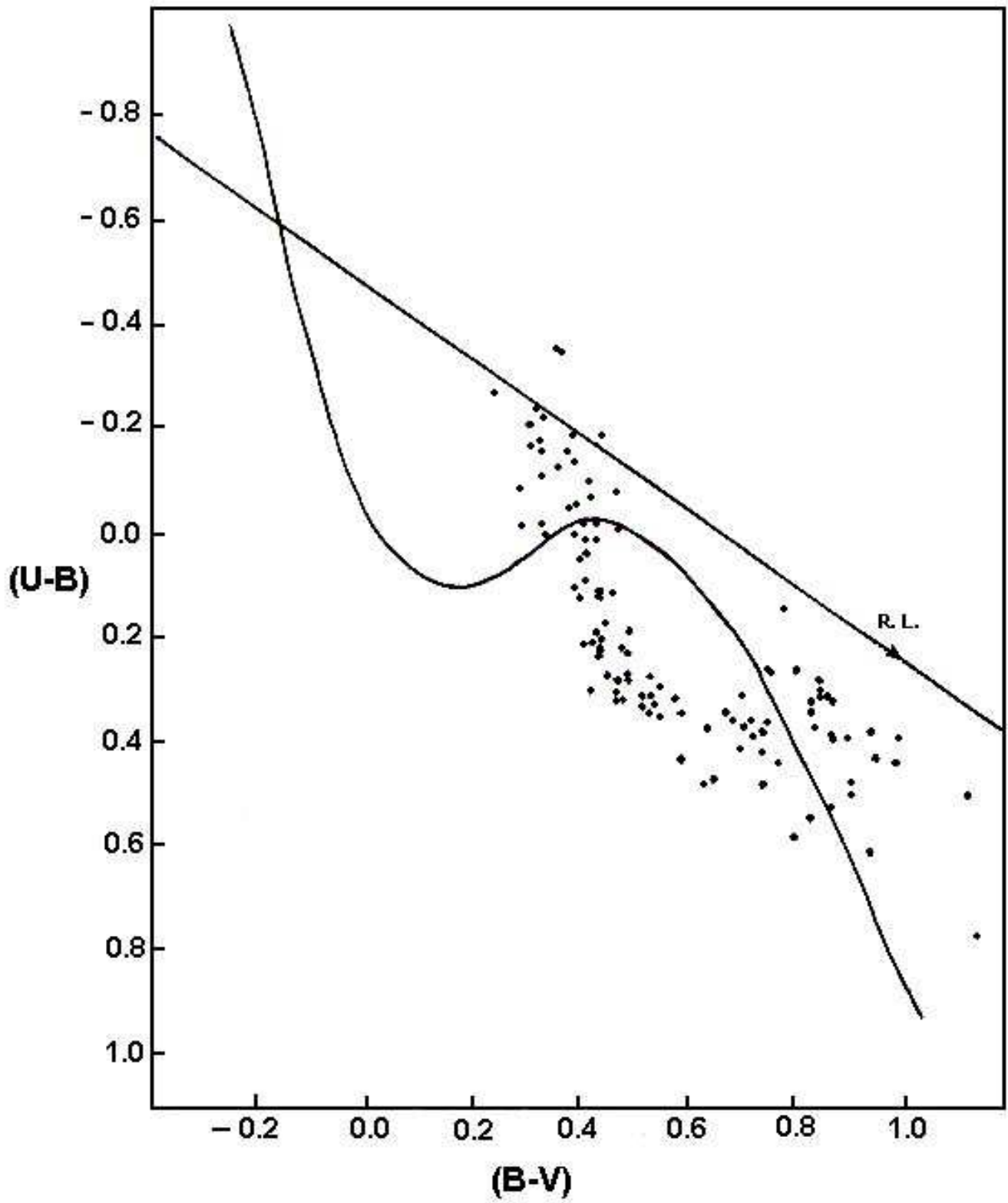}}
\caption{Color-color diagram of M25.}
\end{figure}

\begin{figure}
\centerline{\includegraphics[height=10cm,width=12cm]{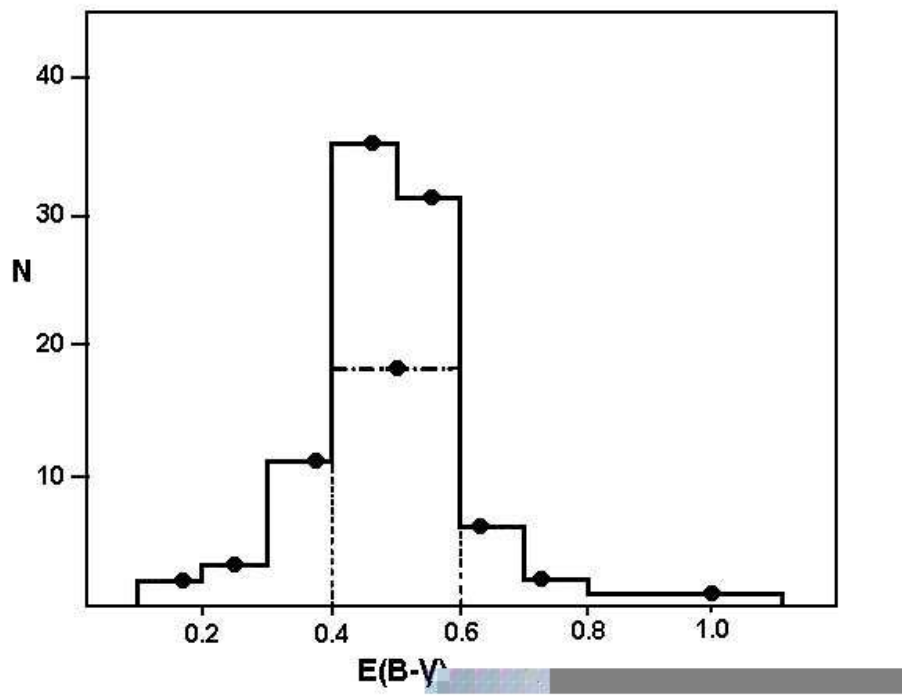}}
\caption{Frequency distribution of color excess for stars in M25.
The solid line represents all members, while the dashed one
represents the members with unambiguous excess.}
\end{figure}

\begin{figure}
\centerline{\includegraphics[height=20cm,width=20cm]{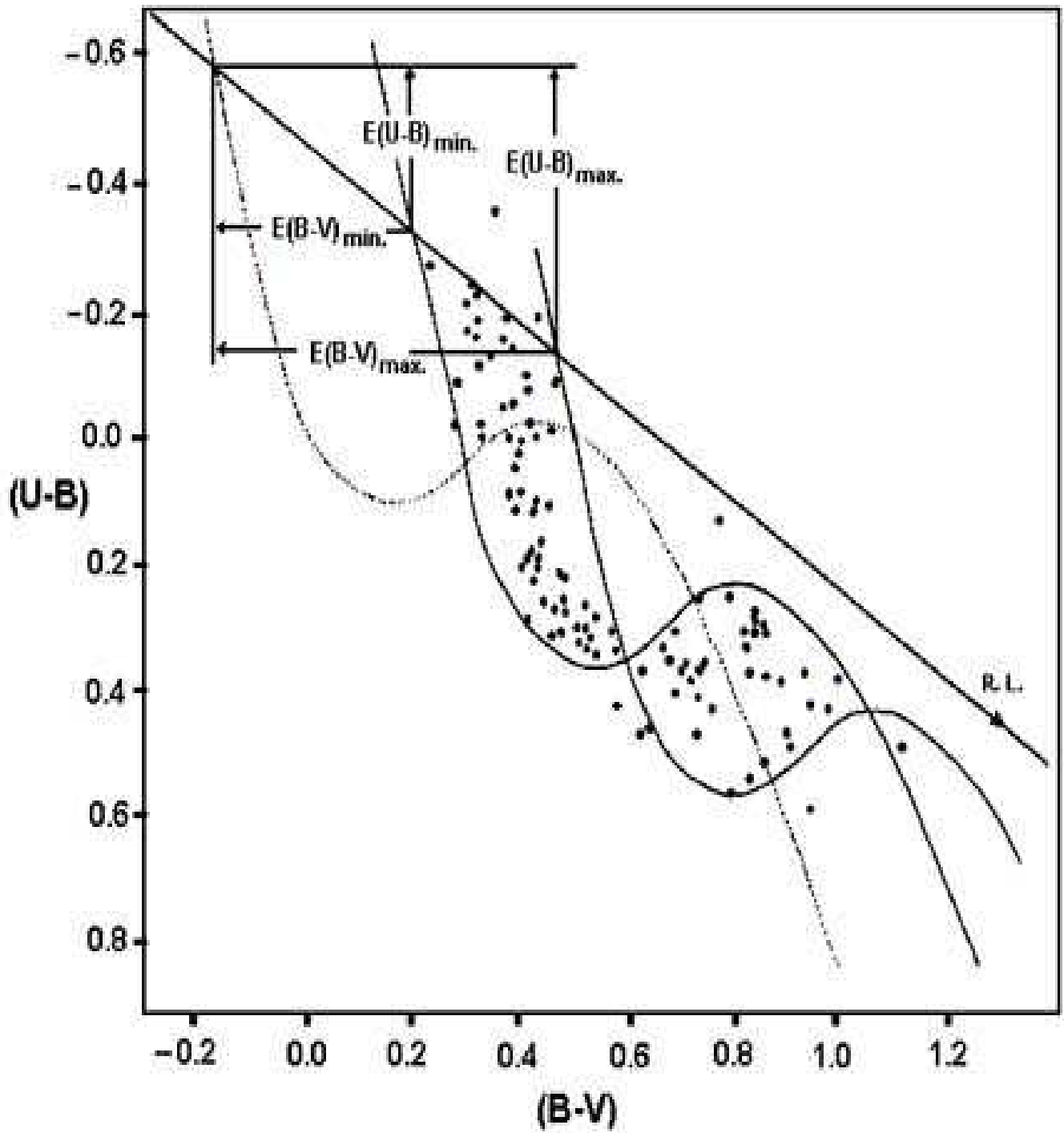}}
\caption{Color-color diagram of M25. The dashed curve is the ZAMS,
while the solid ones represent the maximum and minimum values of
the reddening.}
\end{figure}

\begin{figure}
\centerline{\includegraphics[height=12cm,width=15cm]{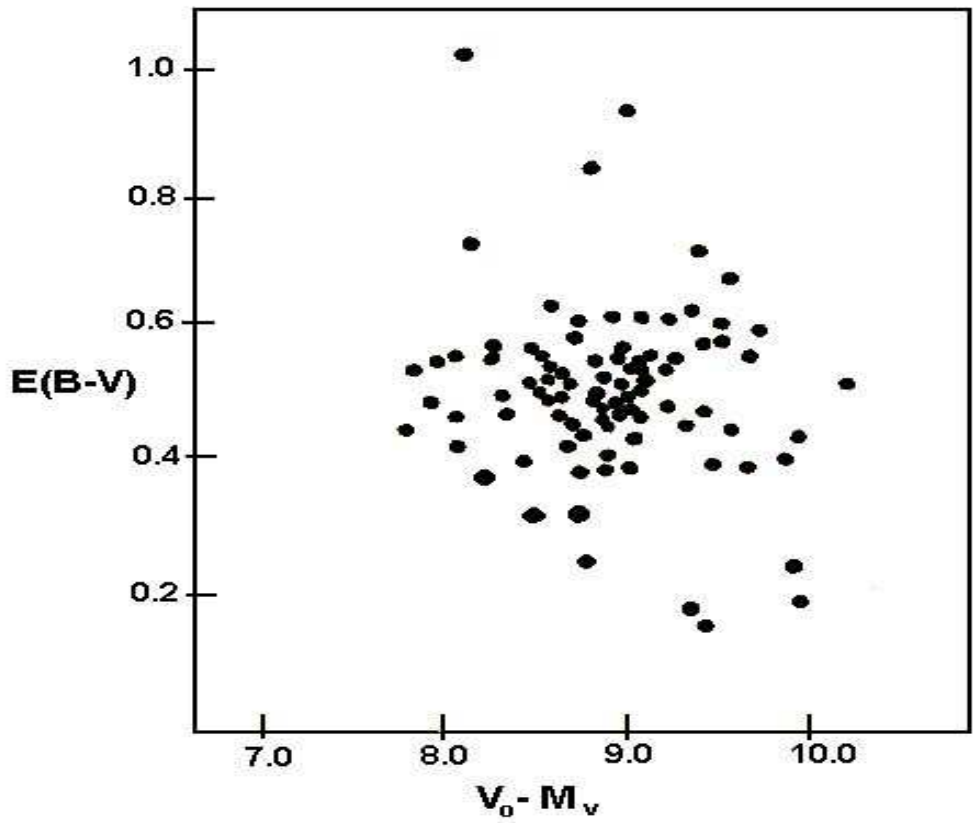}}
\caption{Color excess against distance modulus for each member in
the cluster M25.}
\end{figure}

\begin{figure}
\centerline{\includegraphics[height=10cm,width=15cm]{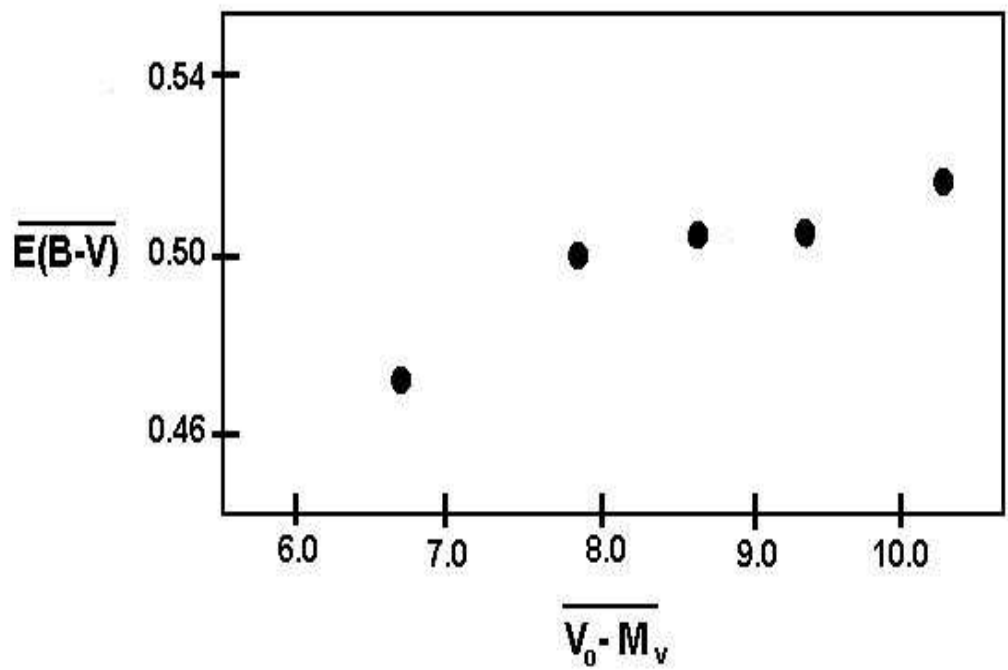}}
\caption{Mean ranges of color excesses against mean ranges of
distance moduluii for members of the cluster.}
\end{figure}

\begin{figure}
\centerline{\includegraphics[height=22cm,width=20cm]{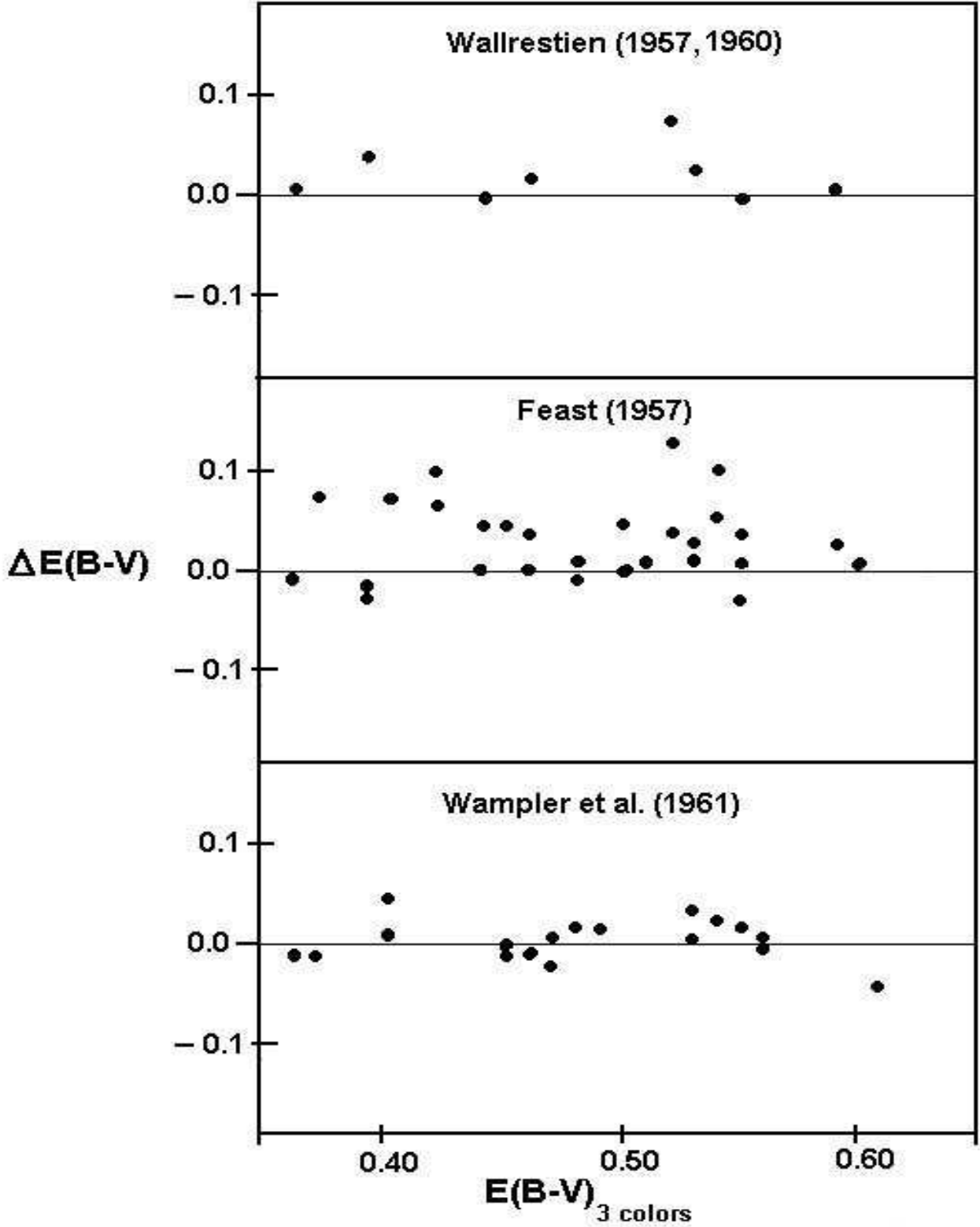}}
\caption{Intercomparison of reddening system in the sense of
3-color photometry minus the corresponding ones of spectral
types.}
\end{figure}

\begin{figure}
\centerline{\includegraphics[height=10cm,width=20cm]{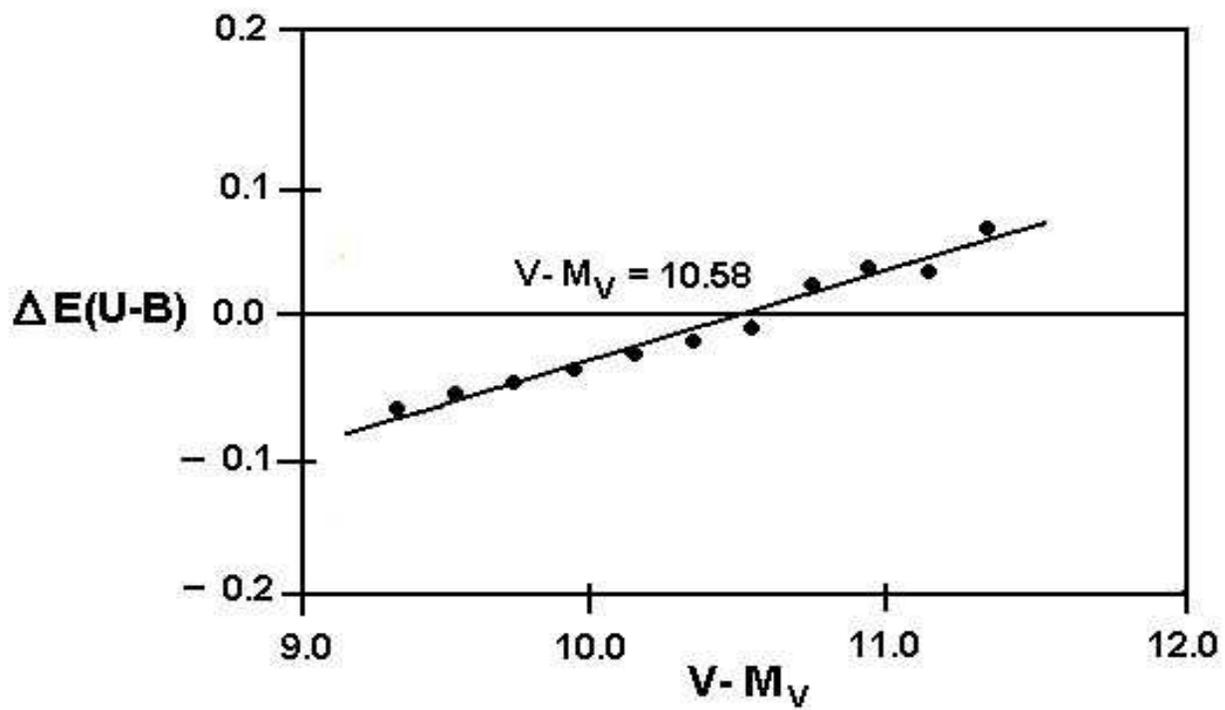}}
\caption{Becker's method fitting to estimate the distance to M25.}
\end{figure}

\begin{figure}
\centerline{\includegraphics[height=22cm,width=18cm]{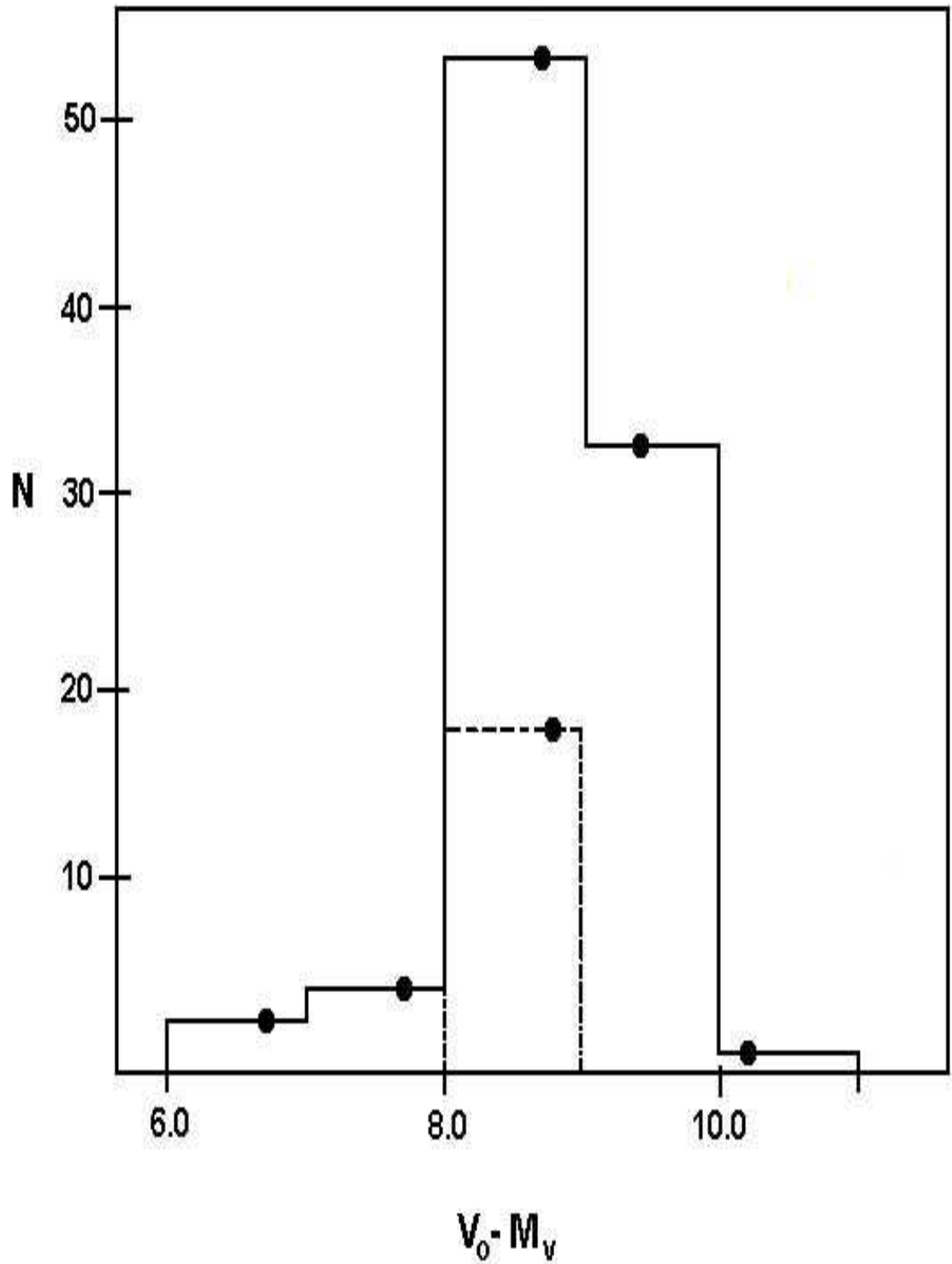}}
\caption{Frequency distribution of distance modulus for stars in
M25. The solid line represents all members, while the dashed one
represents the members with unambiguous excess.}
\end{figure}

\begin{figure}
\centerline{\includegraphics[height=10cm,width=18cm]{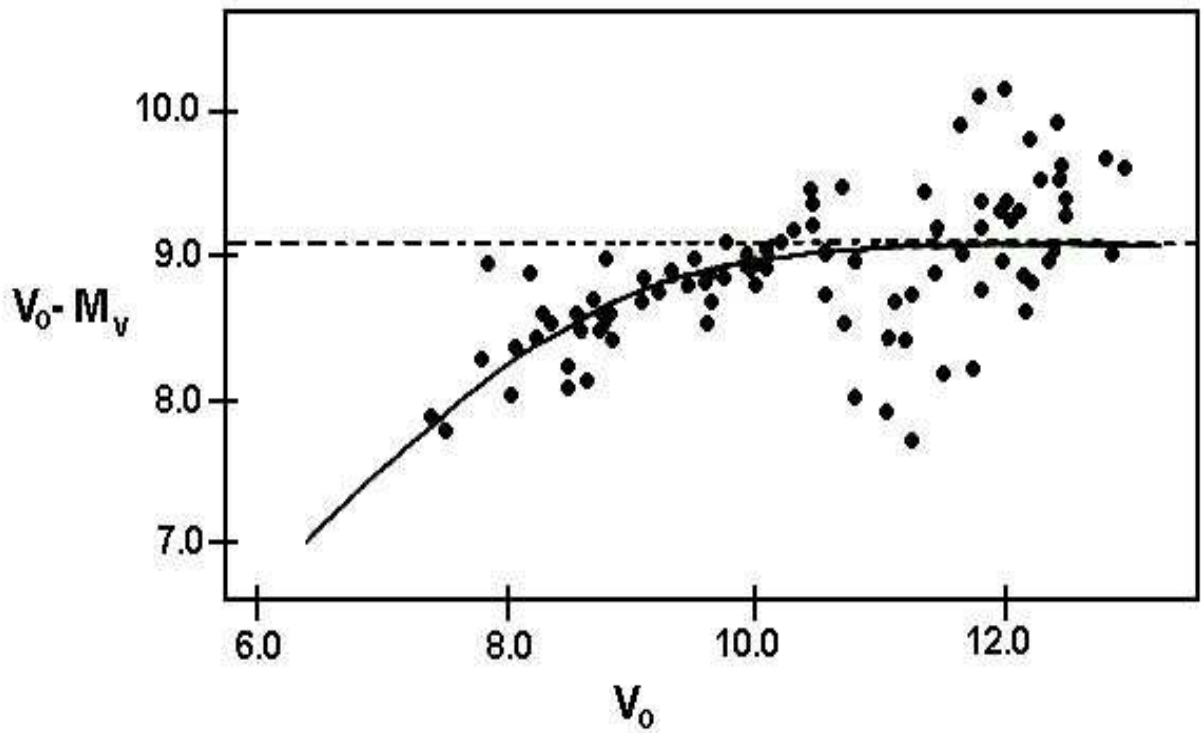}}
\caption{{\it V$_{o}$-(V$_{o}$-M$_{v}$)} diagram for the cluster
members fitted with the mean evolutionary deviation curve of
Johnson (1960)\cite{Joh60a}.}
\end{figure}

\begin{figure}
\centerline{\includegraphics[height=22cm,width=20cm]{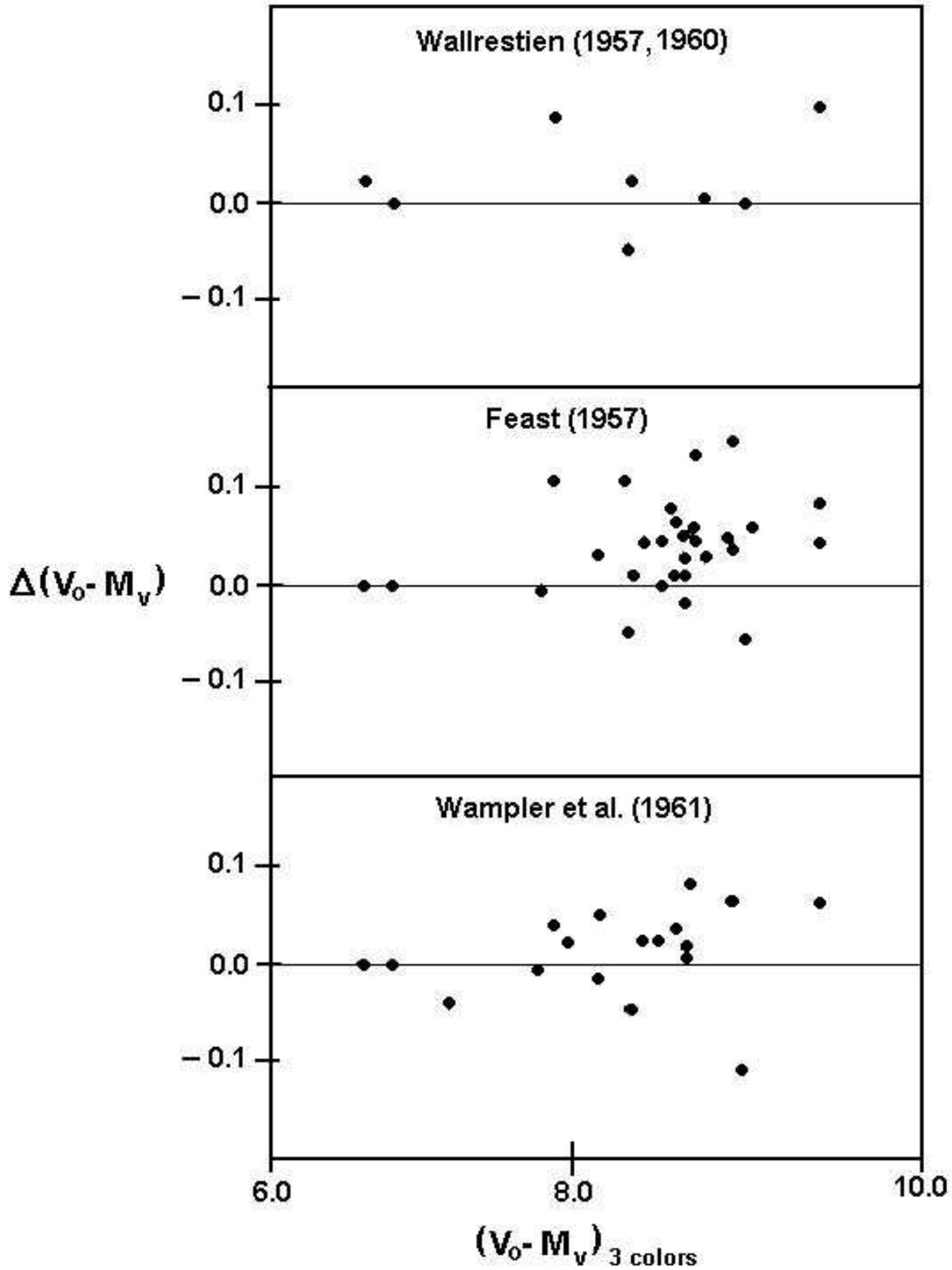}}
\caption{Intercomparison of {\it (V$_{o}$-M$_{v}$)} system in the
sense of 3-color photometry minus the corresponding ones of
spectral types.}
\end{figure}

\begin{figure}
\centerline{\includegraphics[height=22cm,width=20cm]{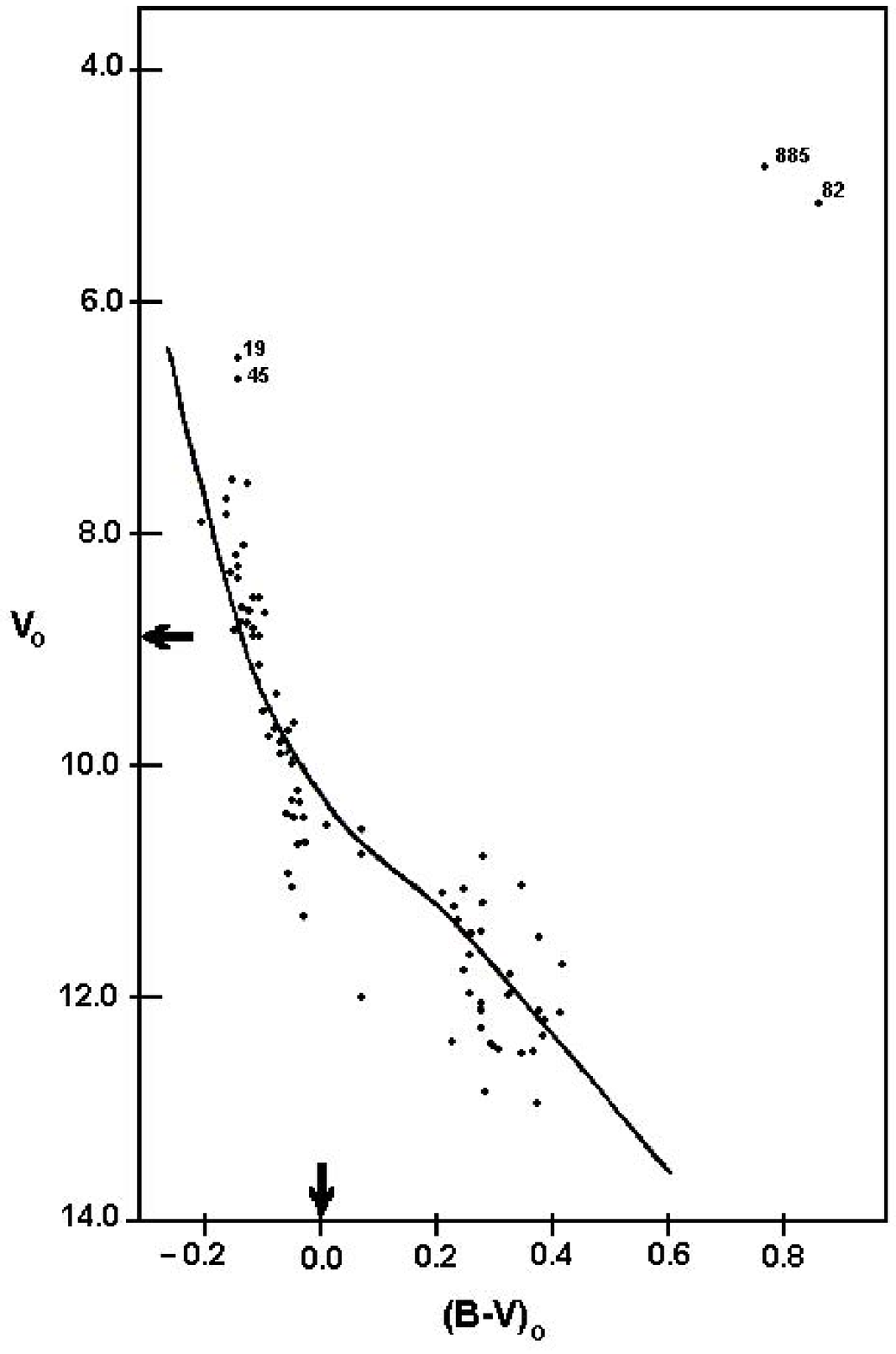}}
\caption{The free {\it V$_{o}$-(B-V)$_{o}$} diagram of M25 fitted
with ZAMS- curve of schmidt-Kaler (1982)\cite{SchK82}. The arrows
show the locations of M$_{v}$=0.0 \& (B-V)$_{o}$=0.0 mag}
\end{figure}

\begin{figure}
\centerline{\includegraphics[height=22cm,width=20cm]{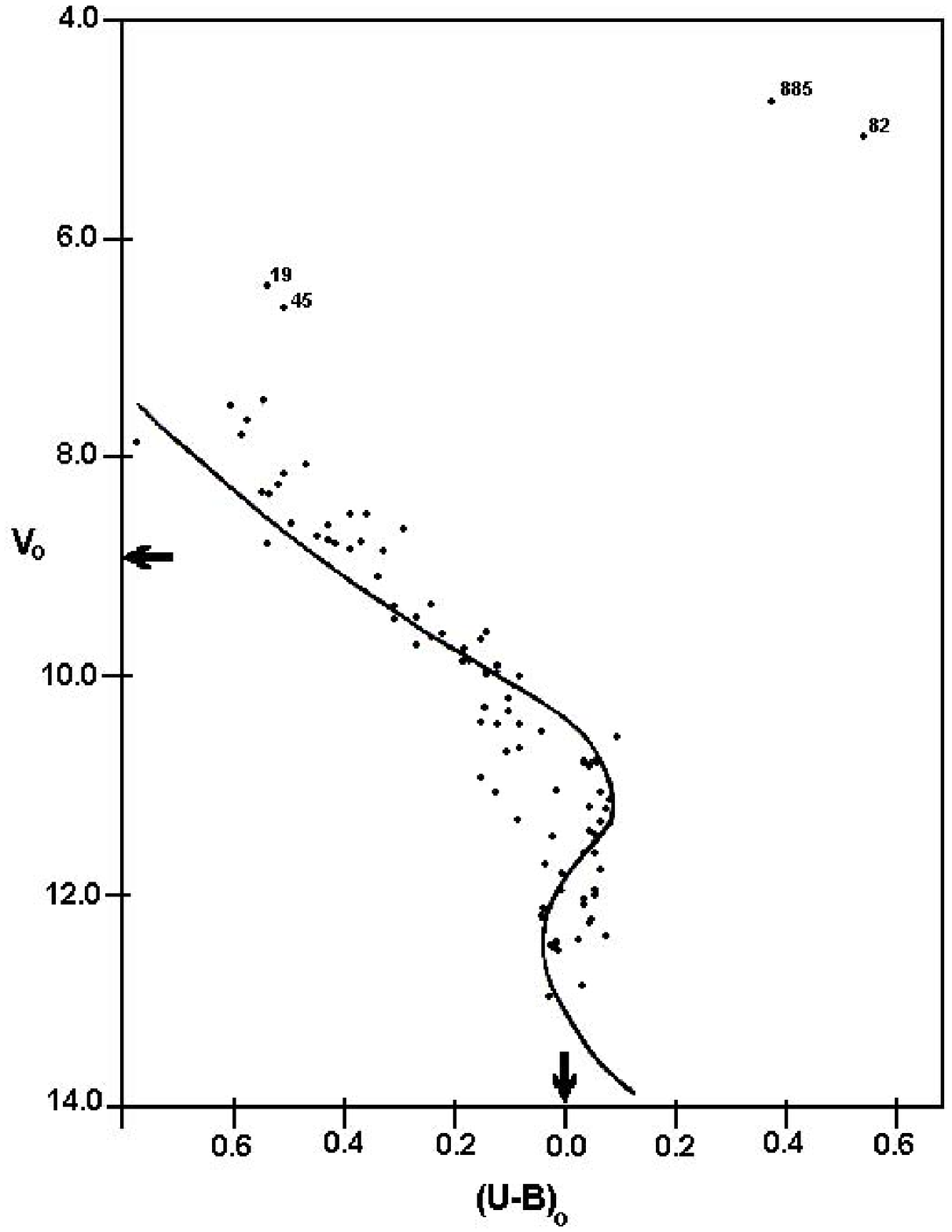}}
\caption{The free {\it V$_{o}$-(U-B)$_{o}$} diagram of M25 fitted
with ZAMS- curve of schmidt-Kaler (1982)\cite{SchK82}. The arrows
show the locations of M$_{v}$=0.0 \& (U-B)$_{o}$=0.0 mag}
\end{figure}

\begin{figure}
\centerline{\includegraphics[height=20cm,width=15cm]{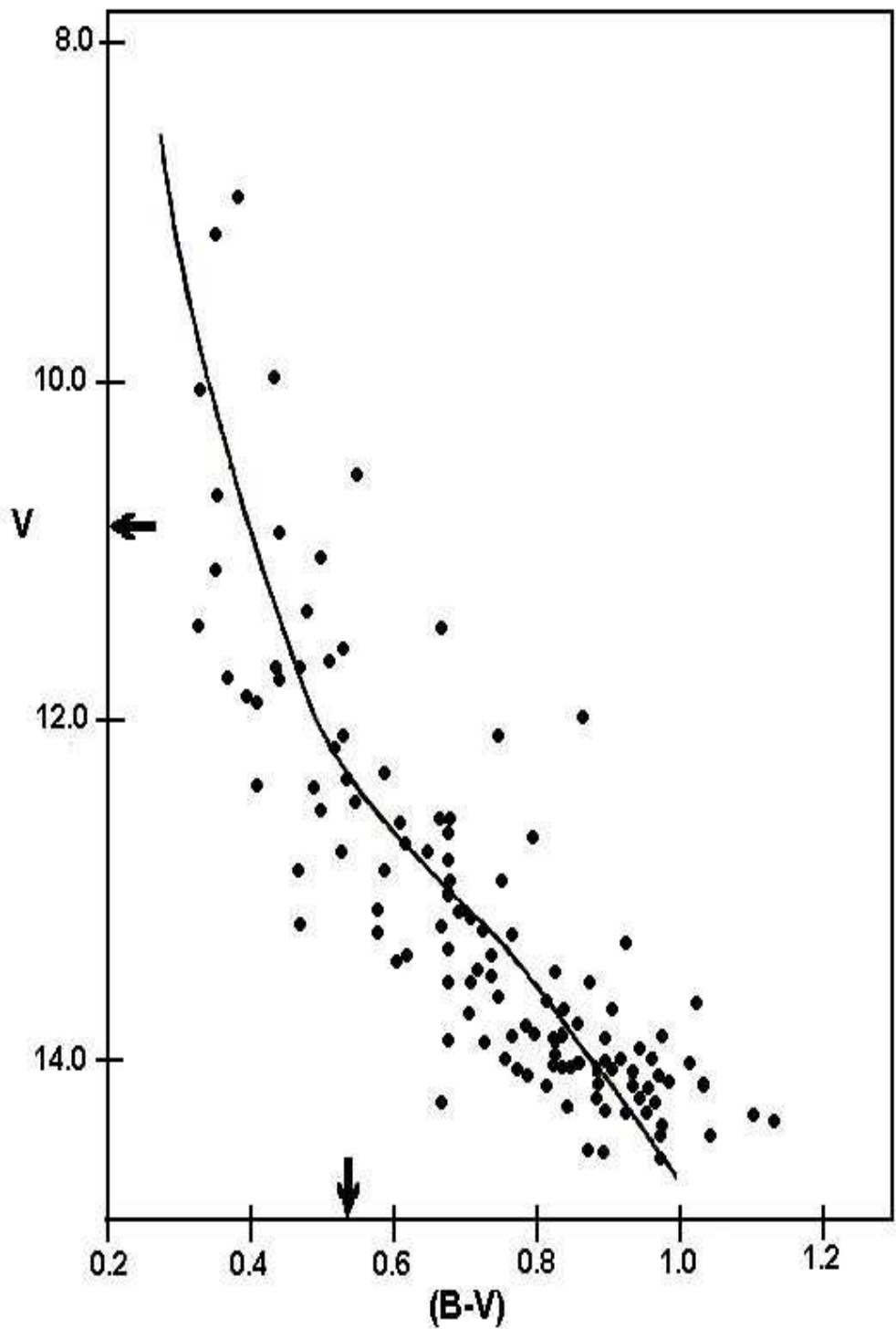}}
\caption{The apparent {\it V-(B-V)} diagram for the stars those
have BV-Observations. The arrows show the locations of M$_{v}$=0.0
\& (B-V)$_{o}$=0.0 mag}
\end{figure}

\begin{figure}
\centerline{\includegraphics[height=18cm,width=18cm]{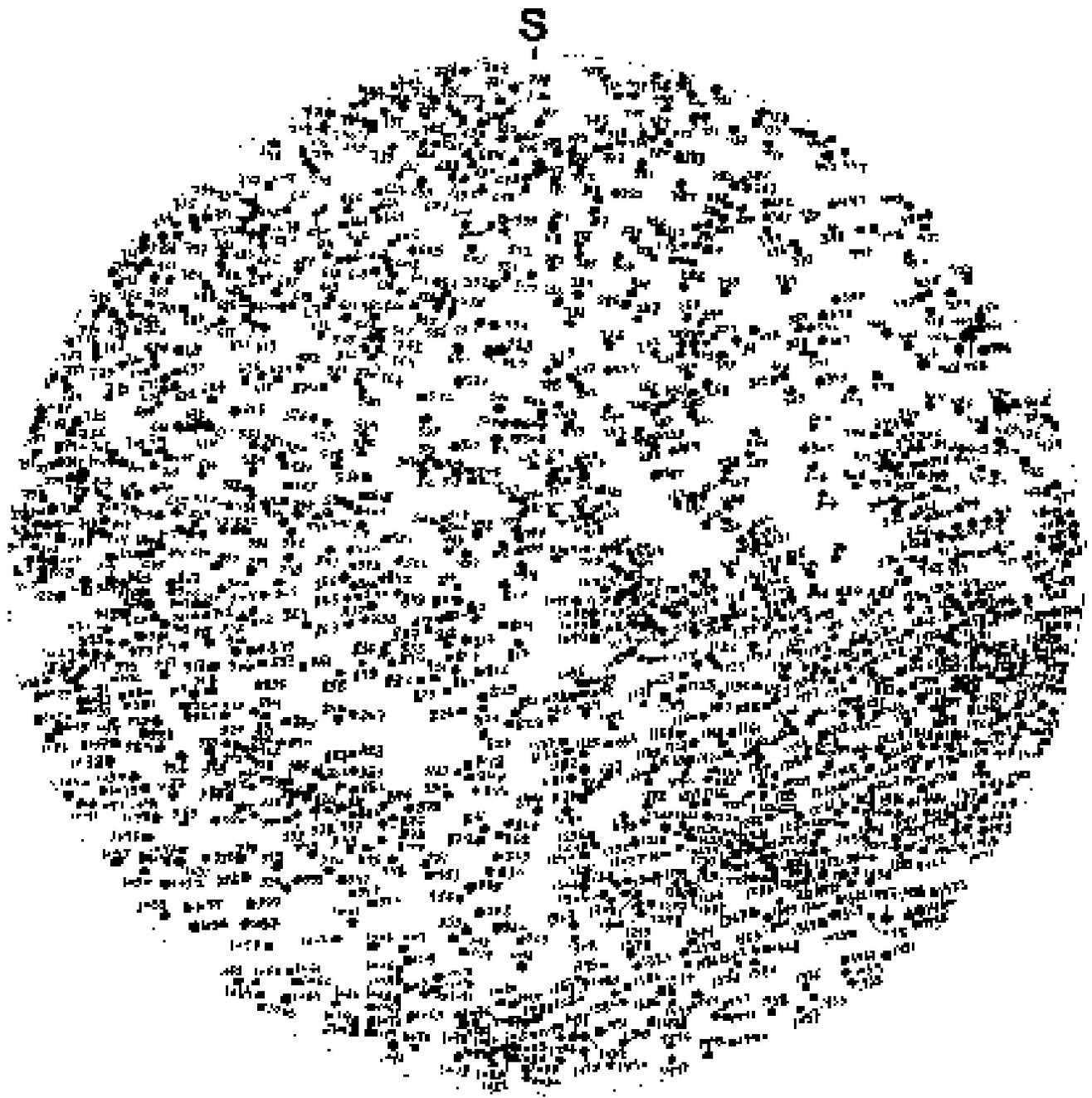}}
\caption{The identification chart for the field of M25.}
\end{figure}

\begin{figure}
\centerline{\includegraphics[height=16cm,width=16cm]{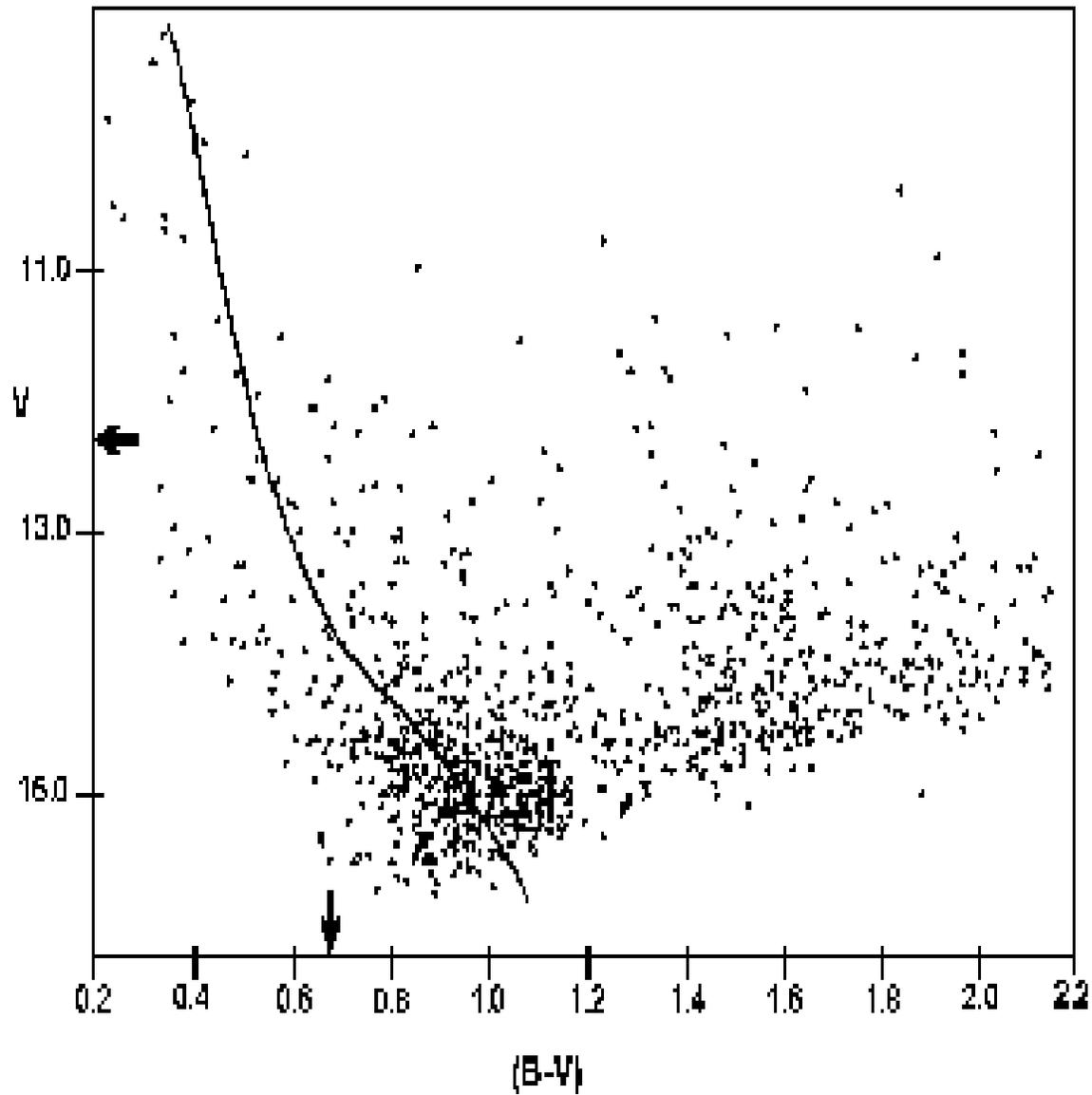}}
\caption{The apparent {\it V-(B-V)} diagram for the field stars of
M25 fitted with ZAMS- curve of schmidt-Kaler (1982)\cite{SchK82}.
The arrows show the locations of M$_{v}$=0.0 \& (B-V)$_{o}$=0.0
mag}
\end{figure}
\end{document}